\documentclass[pra,twocolumn,aps,showpacs]{revtex4}
\usepackage{graphicx}
\usepackage{amsmath}
\usepackage{amsfonts, amssymb, amstext, amsxtra}
\usepackage{longtable}
\usepackage[dvips]{hyperref}

\begin{document}

\title{Designing heterostructures with higher temperature superconductivity}

\author{Karyn Le Hur$^1$}
\author{Chung-Hou Chung$^{1,2}$}
\author{I. Paul$^3$}
\affiliation{$^1$Department of Physics, Yale University, New Haven, CT 06520, USA \\
$^2$Electrophysics Department, National Chiao-Tung University, 
HsinChu, Taiwan, R.O.C. \\
$^3$Institut N\' eel, CNRS/UJF, 25 avenue des Martyrs, BP 166, 38042 Grenoble, France}

\date{\today} 

\begin{abstract}
We propose to increase the superconducting transition temperature 
$T_c$ of strongly correlated materials by designing heterostructures which exhibit a high pairing energy
as a result of magnetic fluctuations. More precisely, applying an effective theory of the doped Mott insulator, we envisage a bilayer Hubbard system where both layers exhibit intrinsic intralayer (intraband) d-wave superconducting correlations. Introducing a finite asymmetry between the hole densities of the two layers such that one layer becomes slightly more underdoped and the other more overdoped, we evidence a visible enhancement of $T_c$ compared to the optimally doped isolated layer. Using the bonding and antibonding band basis, we show that the mechanism behind this enhancement of $T_c$ is the interband pairing correlation mediated by the hole asymmetry which strives to decrease the paramagnetic nodal contribution to the superfluid stiffness. For two identical layers, $T_c$ remains comparable to that of the isolated layer until moderate values of the interlayer single-particle tunneling term. These heterostructures shed new light on fundamental questions related to superconductivity.
\end{abstract}

\maketitle

\section{Introduction}

Since the discovery of high-temperature superconductivity \cite{BM}, considerable efforts have been devoted to finding out how and why it works \cite{Anderson0,Emery,Orenstein,Sachdev,NKP,Fradkin,Anderson,LWN,Leggett,KM}. This puzzling phenomenon --- electrical conduction without resistance at temperatures of up to $\sim 130$ K --- occurs in complex ``copper-oxide'' materials (cuprates). After 1987, the term high-Tc superconductor was used interchangeably with cuprate superconductors until iron-based 
superconductors were discovered \cite{FeAs1,FeAs2}.

The Hubbard model is a well-known model of interacting particles in a lattice, with only two terms in the Hamiltonian: a kinetic term allowing for tunneling (``hopping'') of particles between sites of the lattice and a potential term consisting of an on-site interaction. There are many reasons to believe that the Hubbard model contains most (but maybe not all) of the ingredients necessary for understanding high-temperature superconductivity \cite{ZRsinglet}. At zero hole doping, the single-band Hubbard model definitely captures the insulating behavior of the parent cuprate compounds. The origin of this insulating behavior was described many years ago by Nevill Mott as a correlation effect \cite{Mott} and there is a suppression of the
quasiparticle spectral weight \cite{weight}. In the Mott phase, electron spins form an antiferromagnetic arrangement as a result of the virtual hopping of the antiparallel spins from one copper ion to the next --- the parallel configuration being disallowed by the Pauli exclusion principle. It is relevant to observe that copper-oxide materials are governed by a relatively large magnetic exchange $J\sim 1300$ K which is much larger than the Debye energy of copper.

Upon doping with holes the antiferromagnetism becomes rapidly destroyed and above a certain level superconductivity  occurs with $d_{x^2-y^2}$ pairing symmetry. The $d_{x^2-y^2}$ wave nature of the order parameter has been conclusively shown using phase sensitive experiments \cite{Urbana,Tsuei,Tsuei_review} for example. The earliest experimental observation for d-wave symmetry
is based on the linear decrease of the superfluid stiffness with temperature \cite{Hardy}.
An anisotropic gap with a d-wave order parameter has also been observed through photoemission studies \cite{Shen_review,Shen}. One scenario to explain the $d_{x^2-y^2}$ wave nature of the superconducting gap relies on spin fluctuations at the wavevector $(\pi,\pi)$ which makes the singlet channel attractive at large momentum transfer. This essentially stems from band-structure nesting effects in two dimensions close to half-filling \cite{Scalapino}. A similar pairing occurs in ladder systems as a result of short-range valence bond correlations \cite{Fisher,UKM}. In 1986 it has also been suggested that backscattering from spin fluctuations might lead to the pairing seen in the Bechgaard salts \cite{Eme}. The same year, three papers argued that spin fluctuations are responsible for d-wave pairing in heavy fermion systems \cite{38,39,65}. 

How the electronic structure evolves with doping from a
Mott insulator into a d-wave superconductor is a key issue
in understanding the cuprate phase diagram. Over the years it
has become clear that states in different parts of momentum
space exhibit quite different doping dependences. The
Fermi arcs \cite{Norman,Kanigel} or pockets \cite{Johnson,Taillefer} (near nodal states) retain their coherence as doping is reduced, while the antinodal (near the edge of the first Brillouin zone) states
diminish in coherence, becoming completely
incoherent at strong underdoping. The antinodes open a gap, ``the pseudogap'', which
appears well above the superconducting state. It is important to note that the relationship between the pseudogap and superconductivity is still an open subject \cite{Kondo,Davis} even though some efforts have been accomplished from the theoretical and numerical fronts \cite{KM,John,Yang,Konik,RiceR,Sachdev2,Pruschke,Kotliar,Gull,Civelli,D}. 

There are theoretical indications that the high $T_c$ in the cuprates may result from the large magnetic exchange $J$ \cite{Anderson,Baskaran,Zhang,Paramekanti}. Designing a material that can increase $T_c$ certainly requires a better understanding of the mechanisms that reduce the superfluid stiffness with temperature and with the proximity to the Mott insulating state. A number of recent theoretical \cite{Berg,Okamoto,Altman} and experimental \cite{Bozovic,Yuli} proposals have explored the  possible benefit to combine quite metallic layers with layers of underdoped cuprate materials in heterostructured geometries.

In this paper, we propose to increase $T_c$ using two strongly-correlated Hubbard layers, one layer being slightly underdoped and the other rather overdoped; here, both layers are characterized by prominent d-wave correlations. Using an effective theory of the doped Mott insulator \cite{Anderson,Zhang} for both layers, we report an enhancement of the superconducting transition temperature compared to the optimally doped single layer. Another possibility to increase $T_c$ relies on the presence of a very overdoped ``free electron'' like layer \cite{Altman} (see Fig. 6). 

More precisely, using the bonding and antibonding band representation in the vicinity of optimal doping, we show that the low-energy BCS Hamiltonian exhibits dominant intraband d-wave pairing. Then, we justify how $T_c$  can be enhanced as a result of (interband) additional superconducting fluctuations mediated by the hole asymmetry between the layers. Our results presented in Fig. 1 indeed reveal an enhancement of $T_c$ by $\sim 20\%$  for a slightly underdoped layer with hole density $\sim 0.15$ and an overdoped layer with hole density $\sim 0.25$ --- these include the possible charge redistribution when coupling the layers; optimal doping here refers to hole densities $\sim 0.17$. Our findings may have applications to multilayer materials as well as heterostructures. The analysis performed in this paper uses a purely homogeneous model which does not include stripe or density wave structures \cite{Yuli}. Other Hubbard bilayer systems have been studied in different parametric regimes \cite{Ubbens,Ribeiro}. It is also worth mentioning that  correlated bilayers exhibiting either heavy fermions~\cite{Senthil,Benlagra,Saunders}, composite fermions~\cite{Alicea} or exciton condensates~\cite{Spielman} are also attractive subjects.

The realization of high-Tc superconductivity confined to nanometre-sized interfaces has been a long-standing goal because of potential applications \cite{Ahn}.

The outline of the paper is organized as follows. In Sec. II, we introduce the low-energy theory of the
bilayer system including the effect of Mott physics (large interactions) and we discuss the general methodology. In Sec. III, we address the situation of symmetric layers and show that there is no enhancement of $T_c$; nevertheless, we like to emphasize that the d-wave superconducting state is quite robust toward the proliferation of quasiparticles favored by the single-(quasi)particle tunneling term between the layers and therefore $T_c$ remains almost constant until moderate values of the interlayer tunneling coupling. We also build the BCS Hamiltonian in the band representation of the bilayer system; this is particularly useful to treat the interlayer hopping non-perturbatively. In Sec. IV, we thoroughly compute the superfluid stiffness and $T_c$ in the presence of a finite hole asymmetry.

\begin{figure}[tbh]
\vskip 0.7cm
  \centering
  \includegraphics[width=\linewidth]{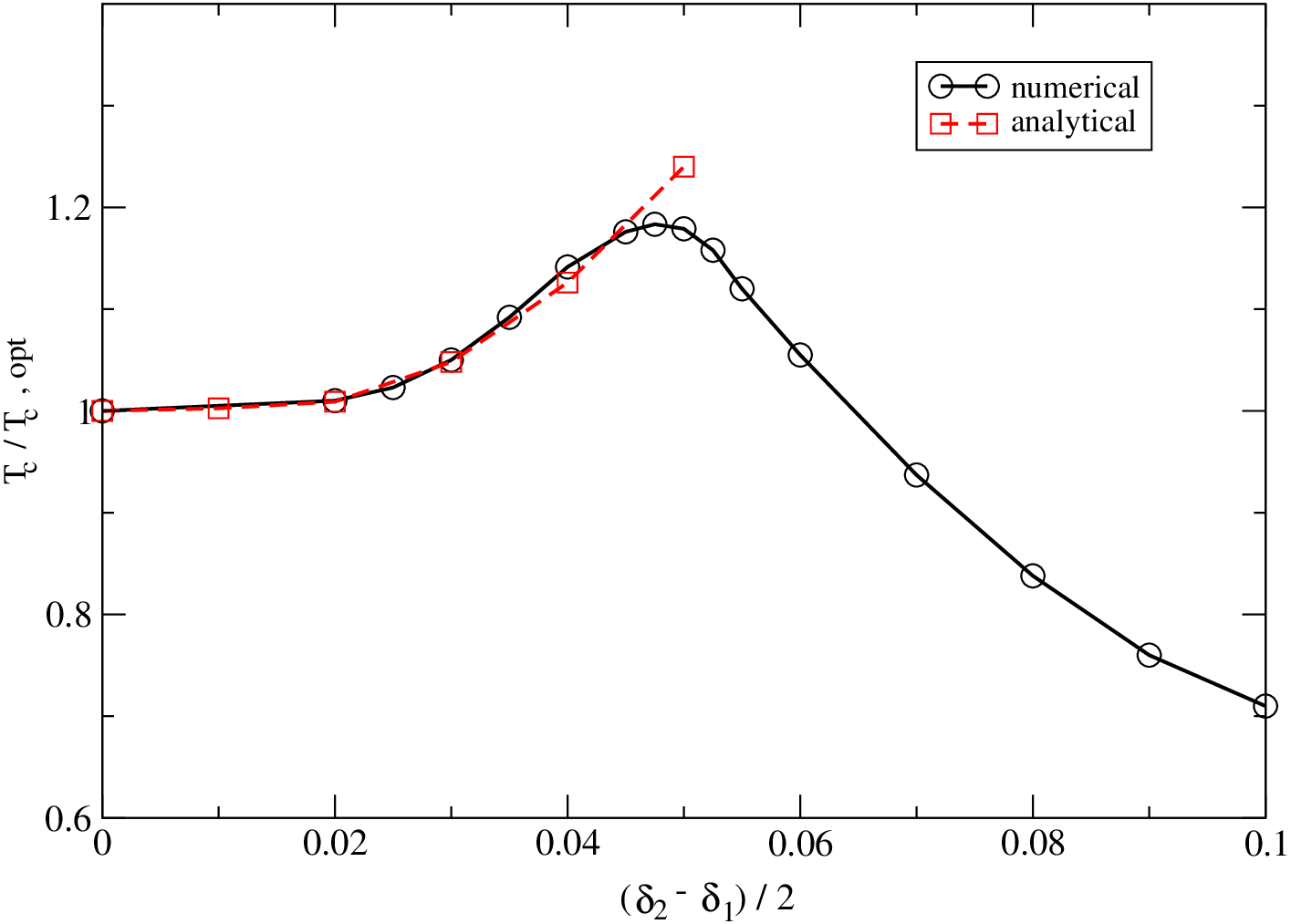}
  \caption{Evolution of $T_c$ for the bilayer when both layers are characterized by prominent intrinsic d-wave correlations; $\delta_1$ and $\delta_2\leq 0.3$ represent the hole densities of the two layers and $(\delta_1+\delta_2)/2$ is fixed to 0.2 close to optimal doping. The red dashed line is obtained from perturbation theory in the band basis. Parameters in the Hamiltonian (1) are $J/t=0.2$, $t'=0$ and $t_{\perp}/t=0.4$. In this figure, $T_{c,opt}$ means  $T_c$ for the bilayer system at optimal doping (see Fig. 5 for $t_{\perp}/t=0.4$).}
  \label{fig:3}
\end{figure}

\section{Model and Methodology}

 Our starting point is the renormalized low-energy theory \cite{Zhang,Gros} which takes into account the proximity of the Mott insulating ground state. Essentially, the Gutzwiller projector \cite{Gutzwiller} ensuring that configurations with
 doubly occupied sites are forbidden is replaced by statistical weighting factors. 
 The projection operator then is eliminated in favor of the reduction factor $g_{ti}=2\delta_i/(1+\delta_i)$ in the kinetic term \cite{Vollhardt} where $\delta_i$ represents the hole doping or the number of holes per site in the layer $i=1$ or $2$. In addition, the projection operator enhances spin-spin correlations in each layer: $g_{si}=4/(1+\delta_i)^2$ \cite{Zhang}. 
 
 The bilayer system in the strong interaction limit then is described by the general Hamiltonian:
\begin{eqnarray}
H &=& -t g_{t1} \sum_{<i,j>\sigma} c^{\dagger}_{i\sigma} c_{j\sigma} -t' g_{t1} \sum_{<<m,n>>\sigma} c^{\dagger}_{m\sigma} c_{n\sigma} +h.c.  \nonumber \\
 &-& t g_{t2} \sum_{<i,j>\sigma} d^{\dagger}_{i\sigma} d_{j\sigma} 
- t' g_{t2} \sum_{<<m,n>>\sigma} d^{\dagger}_{m\sigma} d_{n\sigma} +h.c. \nonumber \\
&-& t_{\perp} g_{t_\perp} \sum_{i\sigma} d^{\dagger}_{i\sigma} c_{i\sigma} + h.c. \nonumber \\
&-& \mu_1 \sum_{i\sigma} c^{\dagger}_{i\sigma} c_{i\sigma} - \mu_2 \sum_{i\sigma} d^{\dagger}_{i\sigma} d_{i\sigma} \nonumber \\
&+& J g_{s1} \sum_{<i,j>} {\bf S}^c_i\cdot {\bf S}^c_j   
+  J g_{s2} \sum_{<i,j>} {\bf S}^d_i\cdot {\bf S}^d_j  \nonumber \\
&+& J_{\perp} g_{s\perp} \sum_{i} {\bf S}^c_i\cdot {\bf S}^d_i,
\end{eqnarray} 
where the operators $c_{\sigma}$ and $d_{\sigma}$ represent electron operators with spin $\sigma$
for layer $1$ and $2$, respectively, $<i,j>$ and $<<m,n>>$ 
refer to nearest-neighbor and next-nearest-neighbor pairs (and we have implicitly assumed $i<j$ and similarly for $m$ and $n$), and ${\bf S}^c$ and ${\bf S}^d$ denote
the spin-1/2 operators in each layer. In our model, the dopings of the two layers are independently tuned through the chemical potentials $\mu_1$ and $\mu_2$. At a general level, the two layers are coupled via the single-particle tunneling contribution $t_{\perp} g_{t_\perp}$ and through the exchange term $J_{\perp} g_{s\perp}$, where $g_{t\perp}=\sqrt{g_{t1} g_{t2}}$ and $g_{s\perp}=\sqrt{g_{s1} g_{s2}}$. 

In Appendix A, we briefly introduce the methodology and the numerical procedure  in the context of the single layer following Zhang and Rice \cite{Zhang}. The advantage of starting with this effective
low-energy theory is that the d-wave superconducting ground state can be studied essentially using 
the usual (unprojected) BCS wavefunction. In the superconducting state, results obtained within this method are in excellent agreement with variational Monte Carlo for projected d-wave states \cite{Paramekanti}.

The main results for the single layer situation are presented in Fig. 2. We check that the Fermi liquid order parameter $\chi_{ij}= (3g_{s}J/4)\sum_{\sigma} \langle c^{\dagger}_{i\sigma} c_{j\sigma}\rangle$ is almost doping-independent whereas the pairing
order parameter $\Delta_{ij}= (3g_{s}J/4) \sum_{\sigma\sigma'}\epsilon_{\sigma\sigma'}\langle c_{i\sigma} c_{j\sigma'} \rangle$ follows the pseudogap (crossover) line \cite{KM} (the indices $i$ and $j$ here involve nearest neighbor sites); we look for mean-field solutions with $\chi_{ij}=\chi$, and $\Delta_{ij}=\Delta$ along $x$-direction and $\Delta_{ij}=-\Delta$ along $y$ direction to ensure d-wave pairing. The d-wave nature
of the order parameter here is dictated by the prominent antiferromagnetic fluctuations at $(\pi,\pi)$ \cite{Scalapino,Baskaran}. Similar to Ref. \cite{Altman} here we assume a unique superconducting gap spreading over the full Fermi surface. In reality, the antinodal points of the Fermi surface are rather governed by the pseudogap \cite{KM,Yang}. In fact, we cannot exclude that the two-gap structure might arise from another competing order with the superconductivity which may alter the results found below. On the other hand, the superfluid density can be formally derived from the quasiparticle contribution close to the nodal points.  Hence, this argument rather supports that $T_c$ mostly depends on the superconducting gap. We have checked that our numerical approach to minimize the free energy perfectly agrees with the mean-field equations (\ref{mfeq_single}).

Within the ``renormalized'' mean-field theory (or equivalently the slave-boson theory \cite{LWN,Ruck,KL,Suz}), the superfluid density at $T=0$ in Eq. (\ref{rhos-singleT0}) is proportional to the hole doping $\delta$, as confirmed experimentally \cite{Uemura}.

The superconducting transition temperature of the isolated layer is evaluated using two complementary approaches \cite{Zhang}. First, the renormalized mean-field theory predicts $T_c\approx g_t \Delta$ (see Appendix B). The second approach consists to evaluate the temperature dependence of the superfluid stiffness. A theory of $T_c$ for the underdoped cuprates has first been built by analogy to the Kosterlitz-Thouless transition in two dimensions. Indeed, Emery and Kivelson in 1995 proposed a model based on phase fluctuations \cite{Emery}. 
In their picture, the pseudogap region is governed by phase fluctuations and at $T_c$ the superfluid stiffness
jumps by the universal amount $2T_c/\pi$. 
\begin{figure}[t]
  \centering
  \vskip 0.6cm
  \includegraphics[width=\linewidth]{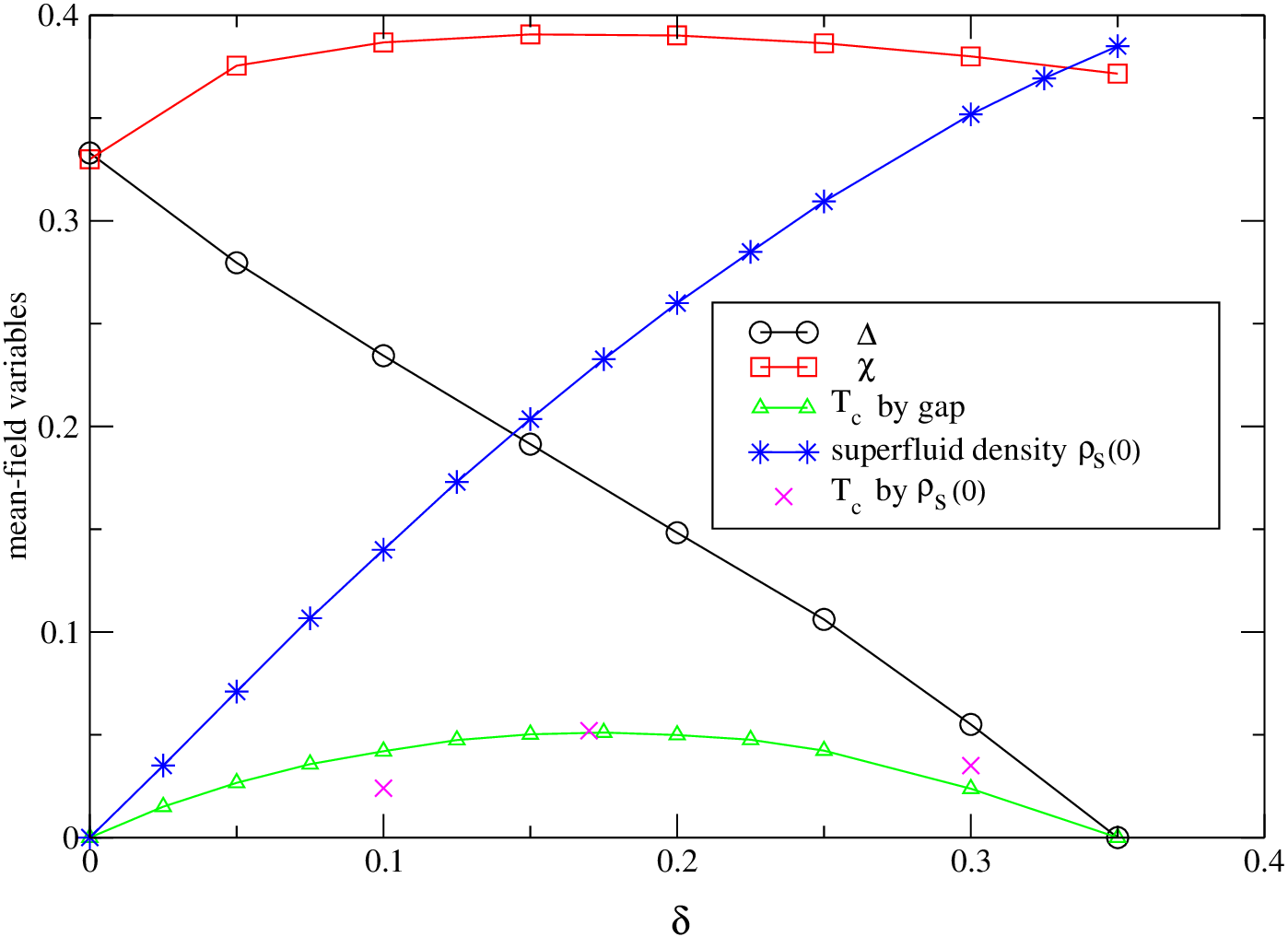}
  \caption{Magnitudes 
of the mean-field variables (in units of $3 g_s J/4$) and the zero-temperature superfluid density 
(in units of $t$) for the single layer described by the renormalized
$t-t'-J$ model versus hole doping. Here, we fix $J/t=0.2$, $t'=0$ 
and $T_c$ is determined from $T_c\approx g_t\Delta$ and from $\rho_s(T_c)=0$.}
  \label{fig:1}
\end{figure}
On the other hand, as mentioned by Lee and Wen in 1997 \cite{LW} the thermal excitation of quasiparticles
near the nodal points rather produce a linear decrease of the superfluid stiffness $\rho_s(T)$ with temperature. This is the earliest experimental evidence of d-wave symmetry \cite{Hardy}. Now, coming back to the Kosterlitz-Thouless scenario, this implies that the $\rho_s$ which controls the transition is not $\rho_s(0)$ but $\rho_s(T)$ which is greatly diminished by quasiparticle excitations, $\rho_s(T)\sim\rho_s(0)-g(T)$ where $g(T)$ is a linear function at low temperatures. Thus, $T_c$ can also be defined by $\rho_s(T_c)=0$. Then, $T_c$ can also be evaluated numerically using Eq. (B1).
Performing an expansion very close to the nodal points leads to \cite{LW, IoffeMillis}
\begin{eqnarray}
&& g(T\rightarrow 0) = a T\nonumber \\
&& a = \alpha^2 \frac{2 \ln 2}{\pi} \frac{v_{F}}{v_{\Delta}},
\label{rhossingle}
\end{eqnarray}
and the ratio between the longitudinal and transverse velocities at the nodes reads (see Appendix B):
\begin{equation}
v_{F}/v_{\Delta} = \frac{2 t g_t+\chi}{\Delta}.
\end{equation}
(We neglect the temperature dependence of $\chi$ and $\Delta$ below $T_c$ as $J\gg T$.) The ratio
$v_F/v_{\Delta}$ is measured through the thermal conductivity \cite{thermal}. An important assumption made in Eq. (\ref{rhossingle}) is that the d-wave quasiparticles are characterized by a {\it renormalized} current \cite{IoffeMillis} $-\alpha e v_F$ (see Appendix B). We introduce the parameter $\alpha$ which can be seen as a phenomenological Landau parameter inherited from the normal state. In principle, the quasiparticle
charge $\alpha e$ should
be determined experimentally \cite{Altman}.

To have a good agreement between the two definitions of $T_c$ and to reproduce the dome-shaped $T_c(\delta)$ phase diagram of the single layer (see Fig. 2), then we fix $\alpha\sim 0.9$. Note that
the value of $\alpha$ depends on the precise scheme used to treat interactions close to the Mott state \cite{Altman}.

\section{Bilayer at Optimal Doping}

First, we apply the methodology of Sec. II to the optimally doped bilayer system. We intend to check
that the superconducting state and therefore $T_c$ are rather stable toward single-(quasi)particle tunneling
favored by the transverse hopping term $t_{\perp}$. In fact, since the transverse hopping term is still weakened by the Gutzwiller statistical weighting factor $g_{t\perp}$ (which is equal to $g_t$ for symmetric layers), we shall show that $T_c$ is almost unchanged until moderate values of $t_{\perp}$ where $g_{t\perp} t_{\perp}\sim J$.  Essentially, the prominent superconducting gap in each layer tends to prevent the proliferation of quasiparticles.

\subsection{Diagonalization in the Band Basis}

In the case of a symmetric bilayer model (with equal hole dopings $\delta_1=\delta_2=\delta$ and $\mu_1=\mu_2=\mu$) it is convenient to use the bonding and antibonding representation: 
\begin{eqnarray}
b_{i\sigma} &=& \frac{1}{\sqrt{2}}(c_{i\sigma}+d_{i\sigma}) \\ \nonumber
a_{i\sigma} &=& \frac{1}{\sqrt{2}}(c_{i\sigma} - d_{i\sigma}),
\end{eqnarray}
which allows to diagonalize all the single-particle hopping terms and therefore to treat $t_{\perp}$ {\it non-perturbatively}.

It is also judicious to introduce explicitly the mean-field order parameters for the two layers:
\begin{eqnarray}
\chi^1_{ij} &=& \frac{3}{4} g_{s} J \sum_{\sigma} \langle c^{\dagger}_{i\sigma} c_{j\sigma} \rangle\nonumber \\
\chi^2_{ij} &=& \frac{3}{4} g_{s} J \sum_{\sigma} \langle d^{\dagger}_{i\sigma} d_{j\sigma} \rangle \nonumber\\
\chi^{\perp}_{ii} &=& \frac{3}{4} g_{s} J_{\perp} \sum_{\sigma} 
\langle c^{\dagger}_{i\sigma} d_{i\sigma} \rangle \nonumber\\
\Delta^1_{ij} &=& \frac{3}{4} g_{s} J \sum_{\sigma\sigma'} \epsilon_{\sigma\sigma'} \langle c_{i\sigma} c_{j\sigma'} \rangle \nonumber \\
\Delta^2_{ij} &=& \frac{3}{4} g_{s} J \sum_{\sigma\sigma'} \epsilon_{\sigma\sigma'} \langle d_{i\sigma} d_{j\sigma'} \rangle \nonumber \\
\Delta^{\perp}_{ii} &=& \frac{3}{4} g_{s} J_{\perp}\sum_{\sigma\sigma'} \epsilon_{\sigma\sigma'} \langle c_{i\sigma} d_{i\sigma'} \rangle.
\end{eqnarray}
For symmetric layers, we can define $g_{si}=g_{s\perp}=g_s=4/(1+\delta)^2$, and we look for mean-field solutions $\chi_{ij}^1=\chi_1$, $\chi_{ij}^2=\chi_2$ where $\chi_1=\chi_2=\chi$, $\chi^{\perp}_{ii}=\chi_{\perp}$, $\Delta^1_{ij}=+\Delta_1$ for two nearest neighbors along $x$ direction and $-\Delta_1$ for two nearest neighbors along $y$ direction, and similarly for the second layer with $\Delta_1=\Delta_2=\Delta$. 

We also check that for $0<t_{\perp}/t<0.5$, the
order parameter $\Delta^{\perp}_{ii}$ is always negligible which means that the only pairing contribution
is the intralayer pure d-wave contribution. (In this paper, we are not interested in the regime of (very) large interlayer transverse hopping amplitudes.) Hereafter, we thus omit the negligible contribution from $\Delta^{\perp}_{ii}$. Therefore the main coupling between the layers is the single-(quasi)particle tunneling term (when assuming $J_{\perp}\propto t_{\perp}^2/U\ll t_{\perp}$) and the term $J_{\perp}$ just renormalizes $t_{\perp}$ by producing a finite $\chi_{\perp}$. 

In the band basis, the mean-field Hamiltonian reads:
\begin{equation}
H_{sym} = H_{Kin} + H_{\Delta} + H_{const},
\end{equation}
where
\begin{eqnarray}
H_{Kin} &=& \sum_{{\bf k}\sigma}\xi_{{\bf k},b} b^{\dagger}_{{\bf k}\sigma} b_{{\bf k}\sigma} + 
        \sum_{{\bf k}\sigma}\xi_{{\bf k},a} a^{\dagger}_{{\bf k}\sigma} a_{{\bf k}\sigma} \nonumber \\
H_{\Delta} &=& \sum_{\bf k} \Delta_{{\bf k},b} b^{\dagger}_{{\bf k}\uparrow} b^{\dagger}_{-{\bf k}\downarrow} 
+\Delta_{{\bf k},a} a^{\dagger}_{{\bf k}\uparrow} a^{\dagger}_{-{\bf k}\downarrow} + h.c.\nonumber \\
H_{const} &=& N_s\sum_{i=1,2,\perp} 
\left[\frac{|\chi_i|^2}{\frac{3}{4} g_{s_i} J} + 
\frac{|\Delta_i|^2}{\frac{3}{4} g_{s_i} J}\right] -2N_s\mu \delta.
\end{eqnarray}
Our convention for the chemical potential follows that of Ref. \cite{Zhang} and $N_s$ is the total number of sites. We identify:
\begin{eqnarray}
\xi_{{\bf k},b/a} &=& -(2tg_t+ \chi) [\cos(k_x)+\cos(k_y)]\nonumber \\
&-&  [4 g_t t' \cos(k_x) \cos(k_y)\pm g_t t_{\perp} \pm \chi_{\perp}] - \mu. 
\end{eqnarray}
For simplicity, the lattice spacing is set to unity and for symmetric layers,
$g_{t\perp}=\sqrt{g_{t1}g_{t2}}=g_t=2\delta/(1+\delta)$.
The Fermi surfaces associated with the two bands get splitted as a result of the
transverse hopping amplitude $t_{\perp}$ and $\chi_{\perp}$. Further,  the pairing parameters $\Delta_{{\bf k},a}$ and
$\Delta_{{\bf k},b}$ are coupled through the mean-field order parameter $\Delta$; more precisely, neglecting 
$\Delta^{\perp}_{ii}$ results in
\begin{equation}
\Delta_{{\bf k},a} = \Delta_{{\bf k},b} = \Delta(\cos(k_x)-\cos(k_y)).
\end{equation}

Interestingly, one can easily diagonalize $H_{sym}$ for any value of $t_{\perp}$ and the mean-field free energy is given by (again $\chi_1=\chi_2=\chi$, $\Delta_1=\Delta_2=\Delta$ and $\Delta_{\perp}=0$):
\begin{eqnarray}
F^{MF}_{sym} &=&  -2 T \sum_{{\bf k},i=a,b}\ln\left[\cosh\left(\frac{E_{{\bf k},i}}{2T}\right)\right] \nonumber \\
&+& N_s\sum_{i=1,2, \perp} 
\left[\frac{|\chi_i|^2}{\frac{3}{4} g_{s_i} J} + 
\frac{|\Delta_i|^2}{\frac{3}{4} g_{s_i} J}\right] -2 N_s\mu \delta.
\end{eqnarray}
The quasi-particle excitation energy for each band is reminiscent of the BCS theory:
\begin{equation}
E_{{\bf k},a/b} = \sqrt{\left(\xi_{{\bf k},a/b}\right)^2 + \left(\Delta_{{\bf k},a/b}\right)^2}.
\end{equation}
The mean-field equations can be obtained by minimizing the free 
energy with respect to $\mu$, $\chi$, $\Delta$ and $\chi_{\perp}$ along the lines of the single layer case and at zero temperature:
\begin{eqnarray}
\delta &=& \frac{1}{2N_s} \sum_{i=a,b} \sum_{{\bf k}} \frac{\xi_{{\bf k},i}}{E_{{\bf k},i}} \\ \nonumber
\chi &=& -\frac{3}{8N_s} g_s J \sum_{i=a,b} \sum_{{\bf k}} (\cos(k_x)+\cos(k_y))\frac{\xi_{{\bf k},i}}{2E_{{\bf k},i}} \\ \nonumber
\Delta &=& \frac{3}{8N_s} g_s J \sum_{i=a,b} \sum_{{\bf k}} (\cos(k_x)-\cos(k_y)) \frac{\Delta_{{\bf k},i}}{2 E_{{\bf k},i}}
\\ \nonumber
\chi_{\perp} &=& \frac{3}{8N_s}g_s J_{\perp} \sum_{j=a,b} \sum_{{\bf k}} j\frac{\xi_{{\bf k},j}}{2 E_{{\bf k},j}},
\end{eqnarray}
where in the last line $j=+$ when $j=a$ and $j=-$ when $j=b$.
Results for the mean-field order parameters at zero temperature are shown in Fig. 3. Here, it is worth mentioning that even though $\chi_{\perp}$ increases as a result of the finite $t_{\perp}$ the d-wave gap remains almost constant reflecting the stability of the d-wave state toward the interlayer single-(quasi)particle tunneling term.

\begin{figure}[t]
  \centering
  \vskip 0.6cm
  \includegraphics[width=8.5cm]{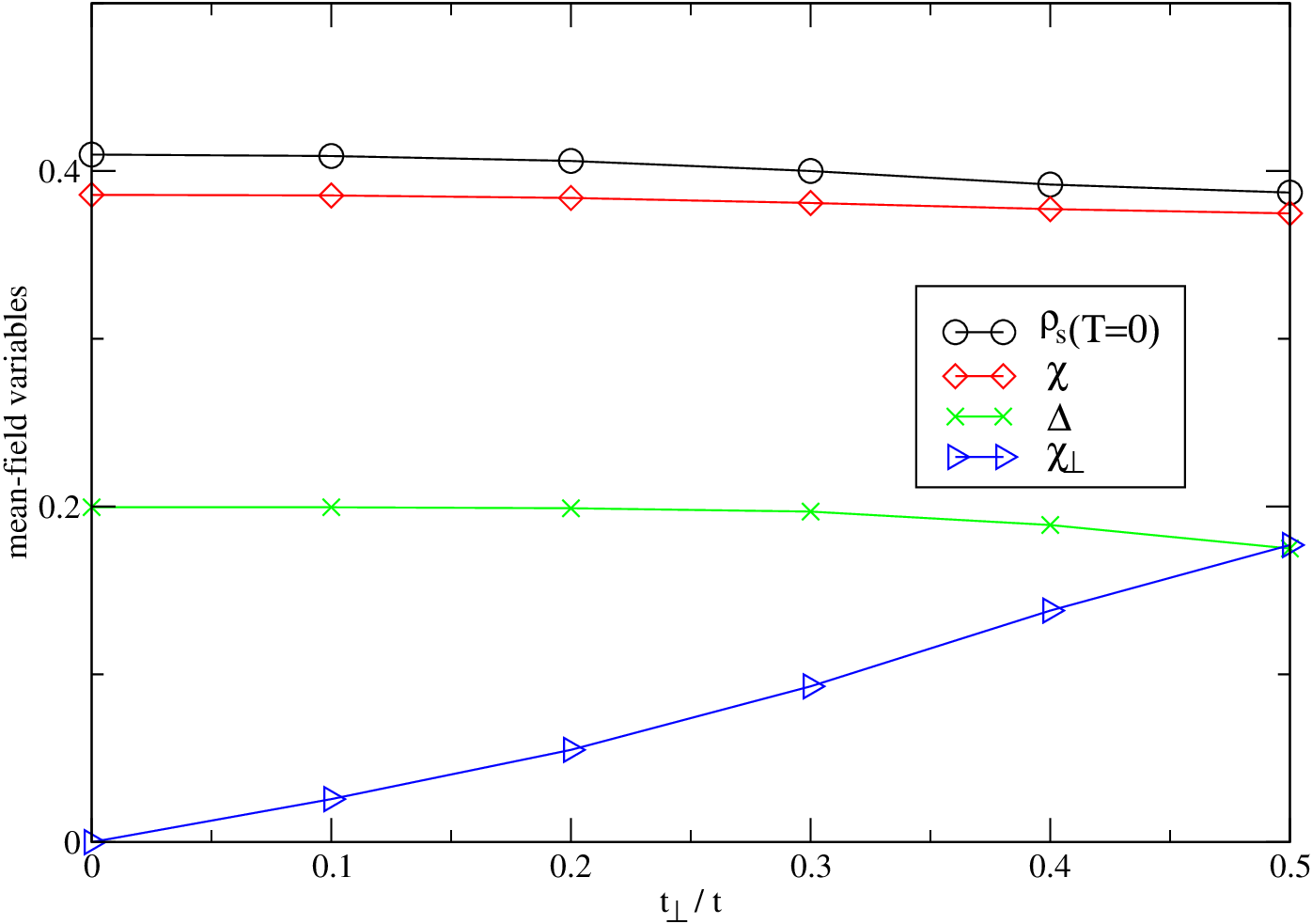}
  \caption{Magnitudes of the order parameters (in units of $3g_sJ/4$) and superfluid density (in units of $t$) for the optimally doped bilayer system at zero temperature as a function of $t_{\perp}$. Optimal doping means $\delta=0.17$, $J/t=0.2$ and for this figure, $t'/t=-0.3\neq 0$.}
  \label{}
\end{figure}

Further, it should be noted that in the band representation the two bands are still characterized by the same d-wave gap. However, details of their distinct Fermi surfaces may matter when evaluating
the superfluid stiffness.

\subsection{Superfluid density and $T_c$}

The diamagnetic current is $({\cal J}_D)_{\mu} = D_{\mu\nu} A_{\nu}$ where
\begin{eqnarray}
D_{\mu\nu} = \sum_{{\bf k}\sigma} \frac{\partial^2 \xi_{\bf k}^0}{\partial k_{\mu} \partial k_{\nu}}\left(c^{\dagger}_{{\bf k}\sigma} c_{{\bf k}\sigma}+d^{\dagger}_{{\bf k}\sigma} d_{{\bf k}\sigma}\right),
\end{eqnarray}
and $\xi_{\bf k}^0$ is the ``bare'' spectrum (the magnetic term $J$ does not contribute to the electric current):
\begin{equation}
\xi_{\bf k}^0 = -2t g_t(\cos(k_x) + \cos(k_y)) - 4 g_t t' \cos(k_x)\cos(k_y).
\end{equation}
The vector potential ${\bf A}$ is directed along the layers. Then, we can rewrite $D_{\mu\nu}$ in the band basis as:
\begin{equation}
D_{\mu\nu} = \sum_{{\bf k}\sigma} \frac{\partial^2 \xi_{\bf k}^0}{\partial k_{\mu} \partial k_{\nu}}\left(b^{\dagger}_{{\bf k}\sigma} b_{{\bf k}\sigma}+a^{\dagger}_{{\bf k}\sigma} a_{{\bf k}\sigma}\right).
\end{equation}
Using the standard BCS ground state wavefunction and the mean-field Hamiltonian in the band representation, then we identify:
\begin{equation}
\sum_{\sigma} \langle b^{\dagger}_{{\bf k}\sigma} b_{{\bf k}\sigma}\rangle = 1-\frac{{\xi}_{{\bf k},b}}{E_{{\bf k},b}}\tanh\left(\frac{E_{{\bf k},b}}{2T}\right).
\end{equation}
We obtain $\langle a^{\dagger}_{{\bf k}\sigma} a_{{\bf k}\sigma}\rangle$ in a similar way.  The zero temperature part of the superfluid density then takes the form:
\begin{equation}
\rho_s(T=0) = \frac{1}{N_s}\sum_{{\bf k},i=a,b} \xi_{{\bf k},0}''\left(1- \frac{{\xi}_{{\bf k},i}}{E_{{\bf k},i}}\right).
\end{equation}
Here,  $\xi_{{\bf k},0}'' = d^2 \xi_{\bf k}^0/d k_x^2$ means that the vector potential is directed along the $x$-axis. For small $t_{\perp}$, we find the following expression:
\begin{eqnarray}
\label{rho_sT0}
\frac{\rho_s(T=0)}{2\rho_s(0)} \approx 1+ \frac{3}{2\rho_s(0)}\sum_{\bf k} 
(t_{\perp}g_{t} +\chi_{\perp})^2\frac{\xi_{\bf k} \xi_{{\bf k},0}''}{E^3_{\bf k}},
\end{eqnarray}
with $\rho_s(0)$ being the zero-temperature superfluid density of the isolated layer and $\xi_{\bf k}$ and $E_{\bf k}$ of the isolated layer are defined in Appendix A. At a general level, 
$\sum_{\bf k}{\xi_{{\bf k},0}''}\xi_{\bf k}/E_{\bf k}^3<0$ which tends to suggest a light downturn of the zero-temperature superfluid density when switching on the interlayer coupling $t_{\perp}$. On the other hand, we check that the exact superfluid density at zero temperature computed from Eq. (B1) does not substantially decrease until quite large values of $t_{\perp}\sim 0.5t$. At small values of $t_{\perp}$, we note an excellent agreement between the exact expression of $\rho_s(T=0)$ in Eq. (B1) and the weak-coupling approximation in Eq. (\ref{rho_sT0}); see Fig. 4. 

\begin{figure}
\centering
\vskip 0.36cm
  \includegraphics[width=8.5cm]{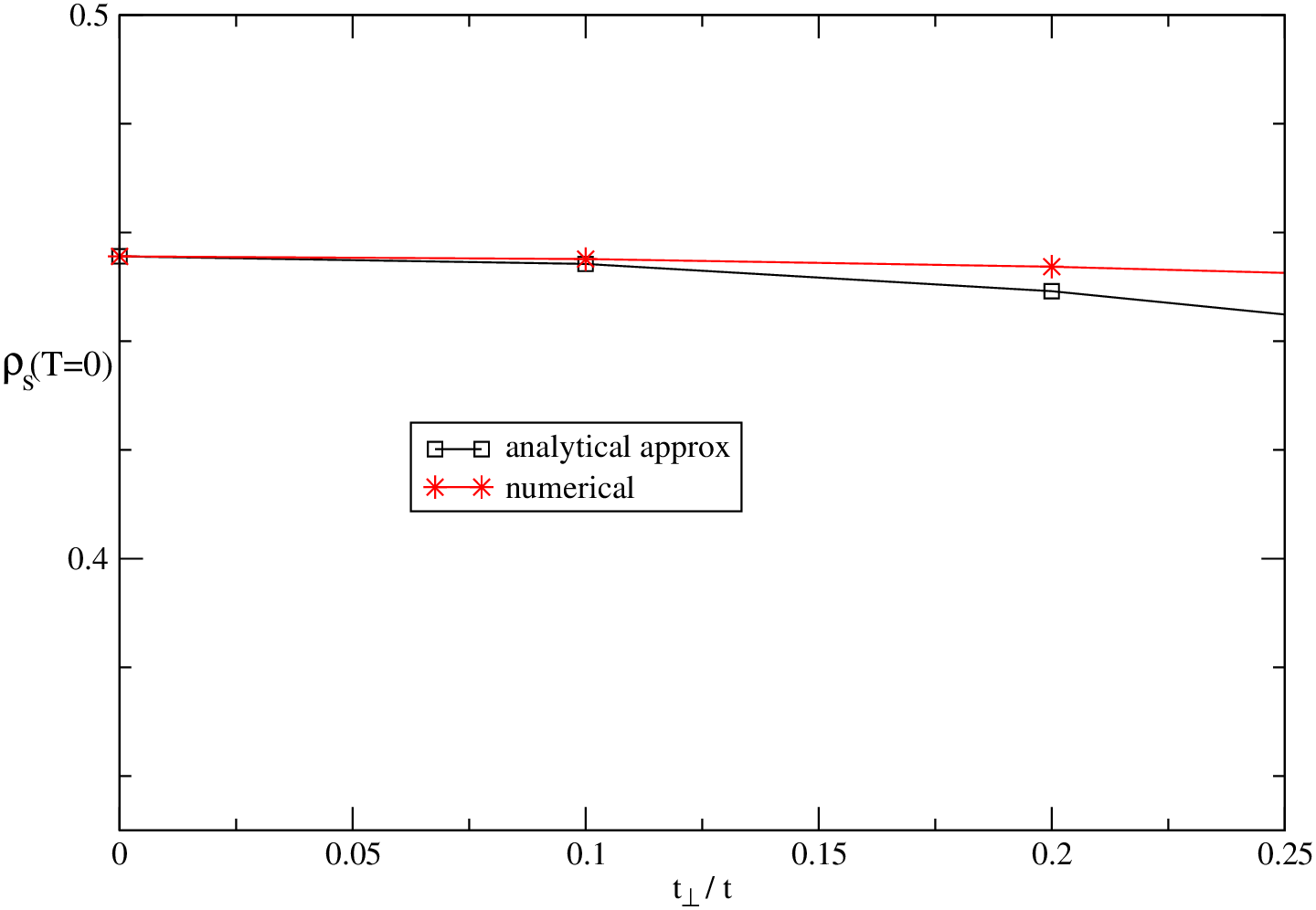}
  \caption{$\rho_s(T=0)$ from Eq. (B1) (in units of $t$) and a weak-coupling approximation in the band basis $(t'=0)$.}
  \label{}
\end{figure}

To compute the critical temperature $T_c$ for the bilayer system at optimal doping, first we follow Goren and Altman \cite{Altman} and diagonalize the mean-field Hamiltonian in the layer basis. The critical temperature $T_c$ is defined via Eq. (B1) by $\rho_s(T_c)=0$ in the layer and band basis. Remember that the transition temperature $T_c$ can be equivalently defined in the band basis using Eq. (10).

At low temperatures, the linear-T dependence of the superfluid stiffness essentially stems from the paramagnetic component (see Appendix B). In the band basis, interestingly, this can be separated into bonding and anti-bonding contributions. Close to the nodal points, we get
\begin{equation}
\rho_s(T) \approx \rho_s(T=0) - \sum_{i=a,b} \alpha_i^2 \frac{2\ln 2}{\pi} T \frac{v_{F,i}}{v_{\Delta,i}},
\label{Tcsym}
\end{equation}
where formally
\begin{equation}
\alpha_i = \alpha \frac{v_F}{v_{F,i}}
\end{equation}
and $v_F$ for the single layer has been defined in Appendix B. It is relevant to mention that in the band basis close to the nodal points the longitudinal and transverse velocities obey:
\begin{equation}
\frac{v_{F,a}}{v_{{\Delta},a}}=\frac{v_{F,b}}{v_{{\Delta},b}}=\frac{2t g_t +\chi}{\Delta}.
\end{equation}
This shows that the ratio $(v_{F,a/b}/v_{{\Delta},a/b})$ remains quite constant until moderate values
of $t_{\perp}$; in particular, it does not involve $\chi_{\perp}$. This allows us to safely approximate 
$\alpha_i\sim\alpha\sim 0.9$.
We shall also mention that since the ratio $v_{F,i}/v_{\Delta,i}$ remains almost identical to that of the isolated layer until moderate values of $t_{\perp}$ this already suggests a very slow reduction of $T_c$ with $t_{\perp}$.

For small values of $t_{\perp}$, the zero-temperature value of the gap obeys:
\begin{equation}
\frac{\Delta(t_{\perp})}{\Delta} = 1-\sum_{\bf k} \frac{(t_{\perp}g_t + \chi_{\perp})^2}{4E_{\bf k}^3}\left(\cos(k_x) -\cos(k_y)\right)^2,
\end{equation}
(where $\Delta$ corresponds to the value of the gap for the isolated layer at optimal doping; see Fig. 2). For small $t_{\perp}$, we predict $T_c=T_{c0}-{\cal C} (g_t t_{\perp}+\chi_{\perp})^2$ with ${\cal C}>0$.

The curve of $T_c$ versus $t_{\perp}$ for the optimally doped bilayer system is presented in Fig. 5.
This unambiguously confirms that the bilayer system at optimal doping is rather stable toward single-(quasi)particle tunneling until quite large values of $t_{\perp}$, defined roughly by $g_t t_{\perp} \sim J$ ($J$ controls the pairing properties of the two layers). The prominent superconducting gap hinders the proliferation of quasiparticles close to the nodal points.

\begin{figure}
  \centering
  \vskip 0.46cm
  \includegraphics[width=\linewidth]{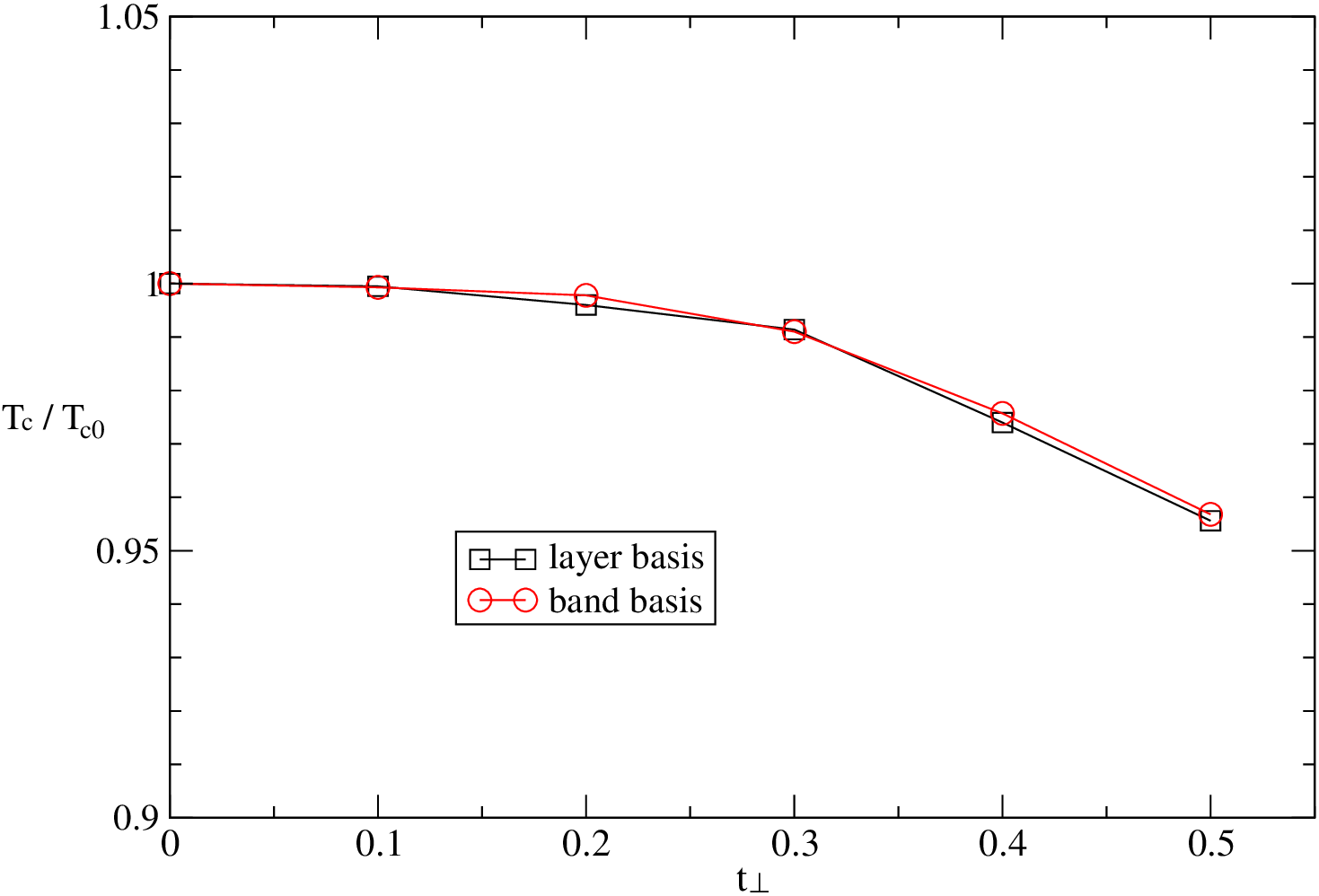}
  \caption{Transition temperature for the bilayer at optimal doping, normalized to $T_{c}$ of the isolated
  layer for the same doping (denoted $T_{c0}$ in this figure), versus $t_{\perp}$. The two curves correspond to the two approaches defined in Sec. III for accessing $T_c$. We use the parameters $t'=0$, $J/t=0.2$ and $\delta=0.17$.}
  \label{}
\end{figure}

\section{Asymmetric layers}

In this Section, we address the case of asymmetrically doped layers, {\it i.e.}, $0\ll \delta_1<0.2$ and $0.2< \delta_2\leq 0.3$ (0.2 is roughly the hole density in each layer at optimal doping). We seek to understand if such a finite asymmetry in the hole dopings of the layers will result in an increase or in a decrease of $T_c$. Using the band basis, we show how a finite hole asymmetry will result in a pairing term coupling the bonding and antibonding bands which helps diminish the quasiparticle nodal contribution for weak asymmetries, then enhancing $T_c$ of the optimally doped situation. The results derived below assume that each layer exhibits  intrinsic d-wave pairing correlations.

\subsection{Interband pairing term}

More precisely, the Hamiltonian becomes
$H=H_{sym}+H_{asy}$ where $H_{sym}$ can be found in Sec. III and
\begin{eqnarray}
H_{asy} &=& \sum_{{\bf k}\sigma} 
\xi_{{\bf k}}^{as} (a_{{\bf k}\sigma}^{\dagger} b_{{\bf k}\sigma} + b_{{\bf k}\sigma}^{\dagger} a_{{\bf k}\sigma})\nonumber \\
 &+&\sum_{\bf k} \Delta_{\bf k}^{as}(a_{{\bf k}\uparrow}^{\dagger} b_{-{\bf k}\downarrow}^{\dagger} + b_{{\bf k}\uparrow}^{\dagger} a_{-{\bf k}\downarrow}^{\dagger} + h.c.).
\end{eqnarray}
In this case, the c-electrons of the first layer and the d-electrons of the second layer are characterized by
different band structures $\xi_{{\bf k},1}$ and $\xi_{{\bf k},2}$ respectively (since $g_{t1}\neq g_{t2}$ and $g_{s1}\neq g_{s2}$). In general, one can decompose
\begin{eqnarray}
\xi_{{\bf k},1} &=& \xi_{\bf k}^{av} + \xi_{\bf k}^{as} \\ \nonumber
\xi_{{\bf k},2} &=& \xi_{\bf k}^{av} - \xi_{\bf k}^{as}.
\end{eqnarray}
Then, we identify:
\begin{eqnarray}
\xi_{{\bf k}}^{as} &=&-\frac{1}{2}(\mu_1-\mu_2) \\ \nonumber
& -& \left(t(g_{t1}-g_{t2}) +\frac{1}{2}(\chi_1-\chi_2)\right)\left(\cos(k_x) + \cos(k_y)\right).
\end{eqnarray}
For simplicity, hereafter we assume that $t'=0$. Similarly, one can rewrite the pairing order parameters
of the two layers as
\begin{eqnarray}
\Delta_{{\bf k},1} &=& \Delta_{\bf k}^{av} + \Delta_{\bf k}^{as} \\ \nonumber
\Delta_{{\bf k},2} &=& \Delta_{\bf k}^{av} - \Delta_{\bf k}^{as},
\end{eqnarray}
which also results in
\begin{eqnarray}
\Delta_{\bf k}^{as} &=& \frac{1}{2}(\Delta_1 -\Delta_2)\left(\cos(k_x)-\cos(k_y)\right).
\end{eqnarray}
Concerning the $H_{sym}$ part this involves $\xi_{\bf k}^{av}$ and $\Delta_{\bf k}^{av}$:
\begin{eqnarray}
\xi_{{\bf k},b/a} &=& -\frac{1}{2}(\mu_1+\mu_2) \mp g_{t\perp}t_{\perp} \mp \chi_{\perp} \\ \nonumber
& -& \left(t(g_{t1}+g_{t2}) +\frac{1}{2}(\chi_1+\chi_2)\right)\left(\cos(k_x) + \cos(k_y)\right)  \\ \nonumber
\Delta_{{\bf k},a} &=& \Delta_{{\bf k},b} = \frac{1}{2}(\Delta_1 + \Delta_2)(\cos(k_x)-\cos(k_y)).
\end{eqnarray}

One can still compute the critical temperature $T_c$ and the superfluid density $\rho_s(T_c)$ (from Eq. (B1)) in the layer basis by diagonalizing the Hamiltonian. On the other hand, to gain some intuition, we also treat the asymmetry terms in the Hamiltonian to second order in perturbation theory (which is justified for quite small asymmetries around the optimal doping). 

Essentially, in the band basis, this results in corrections to $\xi_{{\bf k},a/b}$ and $\Delta_{{\bf k},a/b}$ such that $\xi_{{\bf k},b}$ and $\Delta_{{\bf k},b}$ become
$\tilde{\xi}_{{\bf k},b}$ and $\tilde{\Delta}_{{\bf k},b}$ defined as (see Appendix C):
\begin{eqnarray}
\xi_{{\bf k},b} - \frac{1}{E_{{\bf k},a}^2}\left((\xi_{\bf k}^{as})^2\xi_{{\bf k},a} - (\Delta_{\bf k}^{as})^2\xi_{{\bf k},a} + 2\xi_{\bf k}^{as}\Delta_{\bf k}^{as}\Delta_{{\bf k},a}\right)\hskip 0.3cm && \\ \nonumber
\Delta_{{\bf k},b} - \frac{1}{E_{{\bf k},a}^2}\left((\Delta_{\bf k}^{as})^2\Delta_{{\bf k},a} - (\xi_{\bf k}^{as})^2\Delta_{{\bf k},a} + 2\xi_{\bf k}^{as}\Delta_{\bf k}^{as}\xi_{{\bf k},a}\right), &&
\end{eqnarray}
and similarly for $\tilde{\xi}_{{\bf k},a}$ and $\tilde{\Delta}_{{\bf k},a}$. (There is no first order correction to
the ground state energy.) Note that even though the main pairing contribution is the pure d-wave intraband component, the interband pairing correlations favored by $\Delta^{as}_{\bf k}$ will contribute to reduce the nodal quasiparticle contribution to the superfluid stiffness $\rho_s(T)$. 

From second-order perturbation theory, the effect of $\xi_{\bf k}^{as}$ and $\Delta^{as}_{\bf k}$ is primarily to renormalize the band structure parameters entering into the energies $E_{{\bf k},a}$ and $E_{{\bf k},b}$ of the quasiparticles at the nodal points.

\subsection{Renormalization of nodal contributions}

The next step to compute the transition temperature $T_c$ is to check how $\tilde{\xi}_{{\bf k},a/b}$ and $\tilde{\Delta}_{{\bf k},a/b}$ vary to linear order in k-space for points around the nodal points. 

Notice that the bonding (b) Fermi ``surface''  is larger than the
antibonding (a) Fermi surface. This is a very general fact stemming from the finite interlayer hopping term
$t_{\perp}$. Therefore, we can denote ${\cal N}_b$ and ${\cal N}_a$ the nodes associated
with the bonding and antibonding bands respectively and $k_{ab}>0$ (proportional to $t_{\perp}$) is the k-space distance between the bonding and antibonding Fermi surfaces. Now, $(k_1,k_2)$ are local coordinates with origin
at ${\cal N}_b$ such that:
\begin{eqnarray}
\xi_{{\bf k},b} &=& v_{F,b} k_1 \\ \nonumber
\Delta_{{\bf k},b} &=& v_{\Delta,b}k_2 \\ \nonumber
\xi_{{\bf k},a} &=& v_{F,a} (k_1+k_{ab}) \\ \nonumber
\Delta_{{\bf k},a} &=& v_{\Delta,a}k_2,
\end{eqnarray}
where $k_1, k_2\rightarrow 0$. Then, terms such as
$(\Delta_{{\bf k}}^{as})^2$ are of order $k_2^2$ and therefore do not contribute to linear order. As a result, we can approximate
\begin{eqnarray}
\tilde{\xi}_{{\bf k},b} &\approx& \xi_{{\bf k},b}-\frac{(\xi_{\bf k}^{as})^2\xi_{{\bf k},a}}{E_{{\bf k},a}^2} \\ \nonumber
&\approx& \xi_{{\bf k},b}-\frac{(\xi^{as}_{{\cal N}_B})^2}{v_{F,a} k_{ab}}\left(1-\frac{k_1}{k_{ab}}\right). 
\end{eqnarray}
The first correction term in $\tilde{\xi}_{{\bf k},b}$ shifts
the position of the node ${\cal N}_B$ and the second term gives a correction to $v_{F,b}$:
\begin{equation}
\tilde{v}_{F,b} = v_{F,b} + \frac{(\xi^{as}_{{\cal N}_B})^2}{v_{F,a} k_{ab}^2}.
\end{equation}
Writing $\Delta_{{\bf k}}^{as}=v_{\Delta}^{as}k_2$ we also get:
\begin{equation}
\tilde{\Delta}_{{\bf k},b} \approx \Delta_{{\bf k},b} - \frac{1}{(v_{F,a} k_{ab})^2}\left(2\xi^{as}_{{\cal N}_B}v_{F,a} k_{ab} v_{\Delta}^{as} - (\xi^{as}_{{\cal N}_B})^2 v_{\Delta,a}\right)k_2.
\end{equation}
Therefore, $v_{\Delta,b}$ is renormalized as:
\begin{equation}
\tilde{v}_{\Delta,b} \approx v_{\Delta,b}-\frac{1}{(v_{F,a} k_{ab})^2}\left(2\xi^{as}_{{\cal N}_B}v_{F,a} k_{ab} v_{\Delta}^{as} - (\xi^{as}_{{\cal N}_B})^2 v_{\Delta,a}\right).
\end{equation}
Similarly, we also find
\begin{equation}
\tilde{v}_{F,a} = v_{F,a} + \frac{(\xi^{as}_{{\cal N}_A})^2}{v_{F,b} k_{ab}^2}
\end{equation}
and,
\begin{equation}
\tilde{v}_{\Delta,a} \approx v_{\Delta,a}+\frac{1}{(v_{F,b} k_{ab})^2}\left(2\xi^{as}_{{\cal N}_A}v_{F,b} k_{ab} v_{\Delta}^{as} + (\xi^{as}_{{\cal N}_A})^2 v_{\Delta,b}\right).
\end{equation}
Now, a close inspection of all the contributions leads to:
\begin{eqnarray}
\frac{\tilde{v}_{F,a}}{\tilde{v}_{\Delta,a}} +\frac{\tilde{v}_{F,b}}{\tilde{v}_{\Delta,b}} &\approx& \frac{{v}_{F,a}}{v_{{\Delta},a}}\left(1-\frac{2\xi^{as}_{{\cal N}_A}v_{\Delta}^{as}}{v_{\Delta,a}v_{F,b} k_{ab}}
\right)\\ \nonumber
&+& \frac{{v}_{F,b}}{v_{{\Delta},b}}\left(1+\frac{2\xi^{as}_{{\cal N}_B}v_{\Delta}^{as}}{v_{\Delta,b}v_{F,a} k_{ab}}
\right).
\end{eqnarray}
It should be noted that for very small values of $t_{\perp}$ such that $k_{ab}\rightarrow 0$ then we
could approximate $\xi_{{\cal N}_A}^{as}\sim \xi_{{\cal N}_B}^{as}$ and the nodal corrections
would have practically no effect. For the symmetric case, remember that $v_{F,a}/v_{\Delta,a}=v_{F,b}/v_{\Delta,b}$ and we also check that $v_{\Delta,a}v_{F,b}=v_{\Delta,b}v_{F,a}$.

\subsection{Enhancement of $T_c$}

Now, let assume a moderate value of $t_{\perp}$ such that $0\ll g_{t\perp}t_{\perp}\ll J$. At a general level, we can compute $\xi_{{\cal N}_A}^{as}$ and $\xi_{{\cal N}_B}^{as}$ numerically and then extract the superconducting transition $T_c$. Here, it should be noted that since a prominent $t_{\perp}$ makes 
$\cos(k_x)+\cos(k_y)$ larger for a small Fermi surface and $(g_{t1}-g_{t2})<0$ then this 
immediately implies $\xi_{{\cal N}_A}^{as}>\xi_{{\cal N}_B}^{as}$ (for $\mu_1\sim \mu_2$ and $\chi_1\sim \chi_2$). Based on the nodal quasiparticle contribution only, in the case of a finite (small) hole asymmetry, one then predicts:
\begin{eqnarray}
\label{Tc}
\frac{T_c}{T_{c,opt}} \approx \left(1+\frac{\xi^{as}_{{\cal N}_B}v_{\Delta}^{as}}{v_{\Delta,b}v_{F,a} k_{ab}}-\frac{\xi^{as}_{{\cal N}_A}v_{\Delta}^{as}}{v_{\Delta,a}v_{F,b} k_{ab}}\right)^{-1}>1.
\end{eqnarray}
Here, $T_{c,opt}$ refers to $T_c$ at optimal doping for the symmetric bilayer. In Fig. 1, we compare the  result of Eq. (\ref{Tc}) valid for a small doping asymmetry with the numerical results obtained in the layer basis directly. Both results seem in excellent agreement for small asymmetries. This traduces that
additional (weak) superconducting fluctuations mediated by  $\Delta_{\bf k}^{as}$ mostly affect the nodal contribution(s) whereas other regions (in $k$-space) are protected by the large d-wave gap. 

The enhancement of $T_c$ is attributed to the interplay between interband pairing correlations $(v_{\Delta}^{as})$ and the finite interlayer hopping. This result relies on the existence of a large and small Fermi surface induced by a finite interlayer hopping term. In Fig. 1 we maintain the average hole density $(\delta_1+\delta_2)/2\sim 0.2$ fixed and each layer exhibits intrinsic d-wave pairing correlations (the parameter $\alpha\sim 0.9$ for each layer). In Fig. 6, in contrast, we consider an overdoped layer described by a free-electron like model (with $\alpha_2=1$) and essentially we corroborate the result obtained in Ref. \cite{Altman}, choosing exceptionally $\alpha_1=0.5$ for the underdoped layer. 
Hence, we conclude that two scenarios allow to increase $T_c$ in bilayer systems, one relies on the presence of a very overdoped free-electron layer and the other relies on the enhancement of pairing fluctuations in the band basis induced by a finite hole asymmetry around optimal doping.

\begin{figure}[tbh]
\vskip 0.7cm
  \centering
  \includegraphics[width=\linewidth]{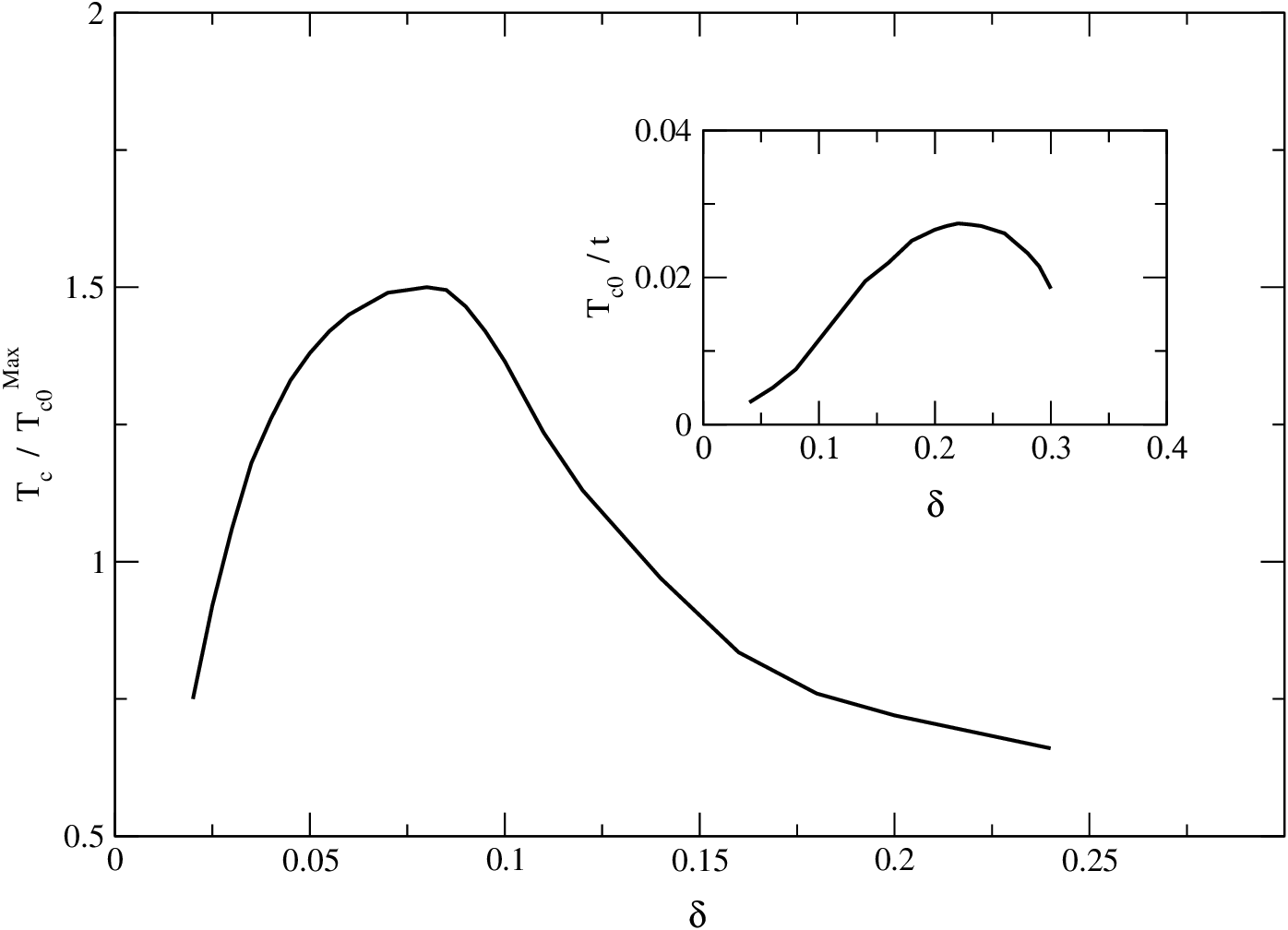}
  \caption{Here, we consider that the overdoped layer is sufficiently doped such that it is described by a free-electron model ($\alpha_2=1$ and $\chi_2=0$) \cite{Altman}; the doping level of the metallic layer  is $\delta_2=0.35$. Parameters are $t_{\perp}/t=0.5$, $J/t=0.3$ and similar to Ref. \cite{Altman}, for this figure, exceptionally we set the renormalized quasiparticle charge of $\alpha_1=0.5$ and $\delta_1=\delta$. Here, $T_{c0}^{Max}$ refers to the maximal value of $T_c$ for the single layer. The enhancement of $T_c$ is in
  agreement with Ref. \cite{Altman} (but it clearly depends on the precise choice of $\alpha_1$.)}
  \label{fig:6}
\end{figure}

\section{Conclusion}

In this paper we corroborate that significant enhancement of $T_c$ in strongly-correlated heterostructures is possible under realistic conditions. More precisely, applying an effective low-energy theory of the doped Mott insulator we have investigated a bilayer Hubbard system where both layers exhibit intrinsic intralayer (intraband) d-wave superconducting correlations. Using the renormalized mean-field theory which is usually well controlled when focusing primarily on the superconducting state, we have evidenced that the increase of $T_c$ results from the delicate balance between the moderate single-particle tunneling term coupling the layers and the finite hole asymmetry around optimal doping which tends to reduce the quasiparticle contribution to the superfluid stiffness by reinforcing superconducting fluctuations. In fact, we have built the BCS Hamiltonian in the band representation of the bilayer system which is particularly judicious to build a non-perturbative theory in the single-particle tunneling term coupling the layers. We have also shown that the d-wave superconducting state is quite robust toward the interlayer single-(quasi)particle tunneling term.
The key point to enhance $T_c$ in these heterostructures is that a finite interlayer hopping produces a larger and smaller Fermi surface and a moderate hole asymmetry between the two layers reinforces (interband) superconducting fluctuations. This scenario requires that both bands are filled ($v_{F,a}\neq 0$ and $v_{F,b}\neq 0$) and hence $t_{\perp}$ should not be too large; a too large $t_{\perp}$ rather favors single quasi-particle interlayer tunneling and therefore is generally not helpful for superconductivity (see Fig. 5).

It is important to distinguish this scenario based on two layers exhibiting prominent d-wave pairing from
the other scenario based on a very overdoped layer described by a metallic bath which essentially 
serves to increase the number of careers \cite{Altman} (see also Fig. 6). The latter situation using a very overdoped metallic layer seems to lead to a more substantial increase of $T_c$ (note that the values of $\alpha_1$ in Figs. 1 and  6 are different).

It should also be mentioned that in this paper we have ignored gauge (phase) fluctuation effects which may play an important role for (quasi-) two-dimensional systems.

We thank T. M. Rice, C. Ahn and F. Walker for stimulating dicussions. This work is supported by DARPA W911NF-10-1-0206, NSF DMR0803200 (KLH), the NSC grant No. 98-2918-I-009-06,
No.98-2112-M-009-010-MY3, the MOE-ATU program,
the NCTS of Taiwan, R.O.C. (CHC). The parallel clusters of the Yale High Performance Computing center provide computational resources. 

\begin{appendix}

\section{Doped Mott insulator and d-wave superconductivity}

The effective Hamiltonian of a doped-Mott insulator takes the form:
\begin{eqnarray}
H_{t-t'-J} &=&
 -t g_t \sum_{<i,j>\sigma} c^{\dagger}_{i\sigma} c_{j\sigma} +h.c. \nonumber \\
&-& t'g_{t} \sum_{<<m,n>>\sigma} c^{\dagger}_{m\sigma} c_{n\sigma} +h.c. \nonumber \\
&-& \mu \sum_{i\sigma} c^{\dagger}_{i\sigma} c_{i\sigma}  
+ J g_{s}\sum_{<i,j>} {\bf S}^c_i \cdot {\bf S}^c_j.
\end{eqnarray}
Here, $<i,j>$ represent the nearest-neighbor pairs and $<<m,n>>$ the next nearest-neighbor pairs (and we implicitly assume that $i<j$ and similarly for $m$ and $n$). The statistical weighting factors for the hopping and spin exchange coupling are 
$g_t=\frac{2\delta}{1+\delta}$ \cite{Vollhardt} and $g_{s} = \frac{4}{(1+\delta)^2}$ \cite{Zhang}, respectively and $\delta$ is the hole doping. Below, we closely follow the notations of Zhang and Rice \cite{Zhang}.

We then introduce the mean-field order parameters:
\begin{eqnarray}
\chi_{ij} &=& \frac{3}{4} g_{s} J \sum_{\sigma} \langle c^{\dagger}_{i\sigma} c_{j\sigma}\rangle,\nonumber \\
\Delta_{ij} &=& \frac{3}{4} g_{s} J\sum_{\sigma\sigma'}\epsilon_{\sigma\sigma'} \langle c_{i\sigma} c_{j\sigma'} \rangle.
\end{eqnarray}
with $\epsilon_{\uparrow\downarrow}=1=-\epsilon_{\downarrow\uparrow}$ and zero otherwise. Then, we look for mean-field solutions with $\chi_{ij}=\chi$, and $\Delta_{ij}=\Delta$ along $x$-direction and $\Delta_{ij}=-\Delta$ along $y$ direction to ensure d-wave pairing. The mean-field Hamiltonian then reads:
\begin{eqnarray}
H^{MF}_{t-t'-J} &=& H_{Kin} + H_{\Delta} + H_{const}\nonumber \\
H_{Kin} &=& \sum_{{\bf k}\sigma}\xi_{{\bf k}} c^{\dagger}_{{\bf k}\sigma} c_{{\bf k}\sigma} \nonumber \\
H_{\Delta} &=& \sum_{\bf k} \frac{\Delta_{{\bf k}}}{2} \left(c^{\dagger}_{{\bf k}\uparrow} c^{\dagger}_{-{\bf k}\downarrow}-
c^{\dagger}_{{\bf k}\downarrow} c^{\dagger}_{-{\bf k}\uparrow}+h.c.\right)\nonumber \\
H_{const} &=& N_s\left[\frac{|\chi|^2}{\frac{3}{4} g_{s} J} + 
\frac{|\Delta|^2}{\frac{3}{4} g_{s} J}\right] - N_s\mu \delta,
\end{eqnarray}
where $N_s$ is the total number of sites (and the chemical potential has been introduced following Zhang and Rice \cite{Zhang}). Assuming a square lattice geometry, we identify:
\begin{eqnarray}
\xi_{\bf{k}} &=& -(2tg_t+ \chi) [\cos(k_x)+\cos(k_y)]\nonumber \\
 &-&  4 g_t t' \cos(k_x) \cos(k_y) - \mu \\ \nonumber
 \Delta_{\bf k} &=& \Delta(\cos(k_x) - \cos(k_y)).
 \end{eqnarray}

The mean-field free energy is given by: 
\begin{eqnarray}
F_{MF} &=& -2 T \sum_{{\bf k}}\ln\left[\cosh\left(\frac{E_{{\bf k}}}{2T}\right)\right] \nonumber\\
&+& 
N_s\left[\frac{|\chi|^2}{\frac{3}{4} g_{s} J} +  \frac{|\Delta|^2}{\frac{3}{4} g_{s} J}\right] - N_s\mu \delta
\label{FMF}
\end{eqnarray}
where the quasi-particle excitation energy obeys
\begin{equation}
E_{\bf{k}} = \sqrt{\xi_{\bf{k}}^2 +\Delta_{\bf{k}}^2}.
\end{equation}
The Boltzmann constant is set to unity.

The mean-field equations can be obtained by directly minimizing the free energy 
with respect to $\chi$, $\Delta$ and by imposing
$(\partial F_{MF}/\partial \mu) = 0$. The mean-field variables of Fig. 2 are solutions of the following equations:
\begin{eqnarray}
\delta &=& \frac{1}{N_s} \sum_{{\bf k}} \frac{\xi_{\bf k}}{E_{\bf k}} \tanh\left(\frac{E_{\bf k}}{2T}\right) 
\nonumber \\
\chi &=&  -\frac{3}{4N_s} g_{s} J \sum_{k} (\cos(k_x) + \cos(k_y)) 
\frac{\xi_{\bf k}}{2 E_{\bf k}} \tanh\left(\frac{E_{\bf k}}{2T}\right),\nonumber \\
\Delta &=& \frac{3}{4N_s} g_{s} J \sum_{\bf k} (\cos(k_x) - \cos(k_y)) 
\frac{\Delta_{\bf k}}{2 E_{\bf k}}\tanh\left(\frac{E_{\bf k}}{2T}\right). \nonumber
\\ 
\label{mfeq_single}
\end{eqnarray}

\section{Superfluid Density and $T_c$}

The superfluid density is formally defined as:
\begin{equation}
\rho_s^{\mu\nu} = \frac{1}{\hbox{Vol}} \left[\frac{\partial^2 F_{MF}}{\partial {\bf A}_{\mu} \partial {\bf A}_{\nu}}\right]_{{\bf A}=0}
\label{rhosdef}
\end{equation}
where $\hbox{Vol}$ is the volume of the system and ${\bf A}$ is the vector potential entering in the kinetic part of the Hamiltonian through a phase:
\begin{eqnarray}
H_{Kin}({\bf A}) &=& -t g_t \sum_{<i,j>\sigma} e^{{\it i} e {\bf A}_{ij}} c^{\dagger}_{i\sigma} c_{j\sigma} \nonumber \\
&-& t'g_{t} \sum_{<<m,n>>\sigma}  e^{{\it i} e {\bf A}_{mn}} c^{\dagger}_{m\sigma} c_{n\sigma}. \nonumber \\ 
\end{eqnarray}
Here, without loss of generality we assume the vector potential to be along the
$x-$axis: ${\bf A} = {\bf A}_x$. 
The superfluid density at any temperature can therefore be evaluated 
numerically through Eq. (\ref{rhosdef}).
 
Further, at $T=0$, the superfluid density can be easily obtained analytically, and it is given by (see Sec. III B):
\begin{equation}
\rho_s(0) = \frac{1}{N_s}\sum_{\bf k} \xi_{{\bf k},0}'' (1-\xi_{\bf k}/E_{\bf k}),
\label{rhos-singleT0}
\end{equation}
where $\xi_{{\bf k},0}'' = d^2\xi^{0}_{{\bf k}}/d k_x^2$ with 
$\xi_{\bf k}^{0} = -2 t g_t(\cos(k_x)+\cos(k_y)) - 4 g_t t' \cos(k_x) \cos(k_y)$ 
being the hopping part of the kinetic energy. [Below, we neglect the T-dependence of this diamagnetic contribution since a power-counting argument shows that this T-dependence is ${\cal O}(T^2)$.]

At finite temperature, the superfluid density is inevitably suppressed by the 
normal state quasiparticle excitations near the four nodal 
points ${\bf q}=(\pm q, \pm q)$  with $q=\pi/2$ at half-filling. In the vicinity of the node $(q, q)$, 
we have the anisotropic Dirac spectrum:
\begin{equation}
E_k \approx \sqrt{v_F^2 k_1^2 + v_{\Delta}^2 k_2^2}
\label{Eknode}
\end{equation}  
where for the square lattice 
\begin{eqnarray}
v_F &=& \sqrt{2} (2 t g_t + \chi)\sin(q) \nonumber\\
v_{\Delta} &=& \sqrt{2} \Delta \sin(q)\nonumber \\  
\cos(q) &=& \frac{-\mu}{2(2 t g_t +\chi)}.
\label{vFvD}
\end{eqnarray}
$v_F$ and $v_{\Delta}$ are the nodal quasiparticle velocities in the longitudinal and transverse directions, 
respectively. For simplicity, here we assume that $t'=0$ allowing a simple analytical solution.
More precisely, by definition 
\begin{eqnarray}
 v_{\Delta} &=&\frac{\partial E_{\bf k}}{\partial k_2}|_{k_1,k_2\rightarrow 0}
\nonumber \\
v_{F} &=&\frac{\partial E_{\bf k}}{\partial k_1}|_{k_1,k_2\rightarrow 0} 
\label{vF}
\end{eqnarray}
where $k_1 = (k_x+k_y - 2 q)/\sqrt{2}$ and $k_2=(k_x-k_y)/\sqrt{2}$ 
with $q$ being associated with the Fermi momentum ${\bf k}_F=(q,q)$ on the 
nodal point ($q = \pi /2$ at half-filling on the square lattice if $t'=0$).
The Fermi momentum obeys:
\begin{equation}
\xi_{{\bf k}_F} = 0 = -(2t g_t +\chi) (\cos(k_x)+\cos(k_y)) - \mu.
\end{equation}
Therefore, we get $\cos(q) = \frac{-\mu}{2(2 t g_t +\chi)}$. We can expand $\xi_{\bf k}$ and $\Delta_{\bf k}$ near a given node,
\begin{eqnarray}
\xi_{\bf k} &\approx& \sqrt{2}(2 t g_t+\chi) \sin(q) k_1 \nonumber \\
\Delta_{\bf k} &\approx& -\sqrt{2} \Delta \sin(q) k_2.
\end{eqnarray}
This results in Eqs. (\ref{Eknode}) and (\ref{vFvD}). 

In the presence of a vector potential, the quasiparticle spectrum exhibits a shift:
\begin{equation}
E(k, {\bf A}) = E(k) - {\bf j}(k) \cdot {\bf A},
\label{EkA}
\end{equation}
where the current ${\bf j}$ carried by the normal state quasi-particles can be formally written
as:
\begin{equation}
{\bf j} = -e \alpha {\bf v}_F.
\end{equation}
Ignoring interactions between the quasiparticles would result in $\alpha = g_t {v}_F^0/{v}_F$ where 
${v}_F^0$ corresponds to the bare Fermi velocity when setting $\chi=0$. Such a choice of $\alpha$ would not allow to reproduce the dome-shaped $T_c(\delta)$. 
Therefore, in this paper, $\alpha\sim 0.9$ will be rather  taken as an effective (constant, doping independent) parameter \cite{IoffeMillis} which can also be regarded as an effective charge. In Eq. (B1), we adjust the vector potential ${\bf A}$ such that it reproduces the correct effective charge.

Using Eqs. (\ref{FMF}), (\ref{rhosdef}), (\ref{Eknode}) and (\ref{EkA}), 
and performing the integral in Eq.(\ref{FMF}) in momentum space near the nodes gives the {\it low-temperature} linear approximation:
\begin{eqnarray}
\rho_s(T) &=& \rho_{s}(T=0)- a T\nonumber \\
a &=& \alpha^2 \frac{2 \ln 2}{\pi} \frac{v_{F}}{v_{\Delta}}.
\label{rhosTsingle}
\end{eqnarray}
The ratio $v_{\Delta}/v_{F}$ near a nodal point  reads:
\begin{equation}
\frac{v_{F}}{v_{\Delta}} = \frac{2 t g_t+\chi}{\Delta}.
\end{equation}

\begin{figure}[t]
\begin{center}
%\vspace{0.47cm}
\includegraphics[width=8.4cm]{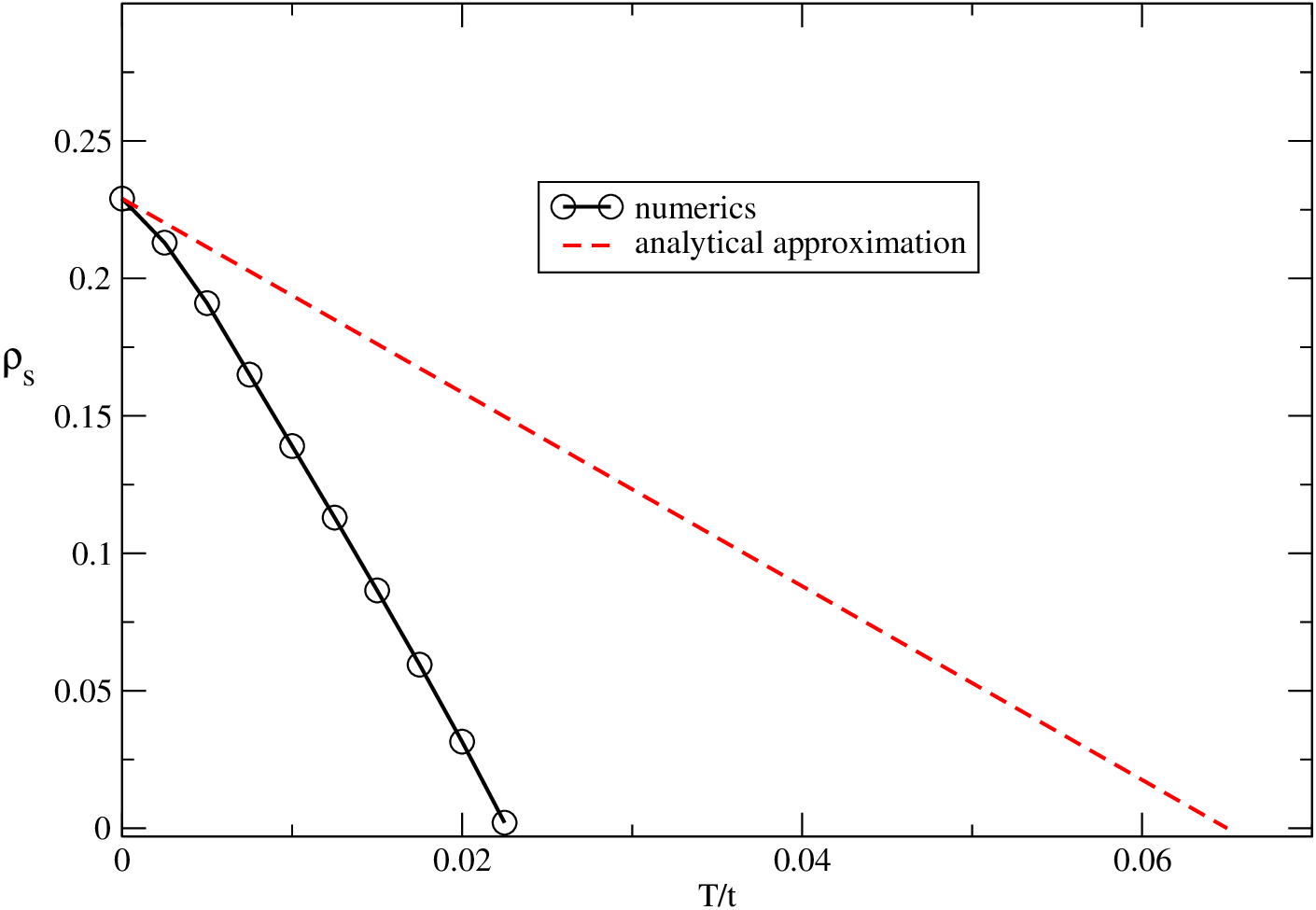}
\label{superf}
\end{center}
\vspace{-0.2cm}
\caption{$\rho_s(T)$ in units of $t$ using Eq. (B1). The analytical approximation (B11) corresponds to a low-temperature expansion very close to the nodes.
The parameters are the same as in Fig. 2 and $\alpha$ is fixed around 
$0.9$. Here, $\delta=0.17$.}
\label{fig2}
\end{figure}

The transition temperature $T_c$ is determined through Eq. (B1) as the temperature at which the superfluid density vanishes due to the thermal excitation of quasiparticles (but, not necessarily nodal); see Fig. 7.

Setting $\alpha\sim 0.9$ in Eq. (B1) through the vector potential ${\bf A}$ allows us to recover a form of $T_c(\delta)$ which is reminiscent of the superconducting order parameter
\begin{equation}
\Delta_{SC}^2 = \langle BCS|P(c^{\dagger}_{i\uparrow} c^{\dagger}_{j\downarrow} c_{i+l\uparrow} c_{j+l\downarrow})P|BCS\rangle\sim g_t^2\Delta^2.
\end{equation}
for a large distance $l$, as shown in Fig. 2.

\section{Perturbation Theory}

Here, we derive Eqs. (29) in the main text obtained by treating $H_{asy}$ in perturbation theory. First, it
is convenient to write the Hamiltonians of each band as $2\times 2$ block matrices. When $H_{asym}=0$ then this results in
\begin{eqnarray}
H_B &=& \left[
  \begin{array}{ c c }
     \xi_{b}  & \hskip 0.2cm \Delta_{b}\\
    \Delta_{b}  &\hskip 0.2cm  -\xi_{b}
  \end{array} \right] \\ \nonumber
H_A &=& \left[
  \begin{array}{ c c }
     \xi_{a}  & \hskip 0.2cm  \Delta_{a}\\
    \Delta_{a}  &\hskip 0.2cm  -\xi_{a}
  \end{array} \right].
  \end{eqnarray}
  For simplicity, we suppress the momentum index ${\bf k}$. The mixing between the bonding and antibonding sectors is given by $V_{AB}=V_{BA}=H_{asy}$:
\begin{eqnarray}
H_{asy}=\left[
  \begin{array}{ c c }
     \xi_{as}  & \hskip 0.2cm \Delta_{as}\\
    \Delta_{as}  &\hskip 0.2cm  -\xi_{as}
  \end{array} \right].
  \end{eqnarray}
  Solving the time-independent Schr\" odinger equation and integrating out the A-subsystem the effective
  Hamiltonian for B subsystem is given by:
  \begin{equation}
  \tilde{H}_B = H_B +V_{BA}(-H_A)^{-1} V_{AB}
 \end{equation}
 with $V_{BA}(-H_A)^{-1} V_{AB}=$
 \begin{eqnarray}
 \frac{1}{E_a^2}
 \left[
  \begin{array}{ c c }
  \xi_{as}  & \hskip 0.2cm \Delta_{as}\\
    \Delta_{as}  &\hskip 0.2cm  -\xi_{as}
  \end{array} \right]  
  \left[
    \begin{array}{ c c }
     -\xi_{a}  & \hskip 0.2cm  -\Delta_{a}\\
    -\Delta_{a}  &\hskip 0.2cm  \xi_{a}
  \end{array} \right]
  \left[
  \begin{array}{ c c }
     \xi_{as}  & \hskip 0.2cm \Delta_{as}\\
    \Delta_{as}  &\hskip 0.2cm  -\xi_{as}
  \end{array} \right].  
  \end{eqnarray}
  Therefore, we check that
   \begin{eqnarray}
 H_B+V_{BA}(-H_A)^{-1} V_{AB}=
 \left[
  \begin{array}{ c c }
   \tilde{\xi}_b & \hskip 0.2cm \tilde{\Delta}_b \\
    \tilde{\Delta}_b &\hskip 0.2cm  -\tilde{\xi}_b
  \end{array} \right],
  \end{eqnarray}  
  where $\tilde{\xi}_{b}$ and $\tilde{\Delta}_{b}$ are precisely defined as
\begin{eqnarray}
\tilde{\xi}_b &=& \xi_{b} - \frac{1}{E_{a}^2}\left((\xi^{as})^2\xi_{a} - (\Delta^{as})^2\xi_{a} + 2\xi^{as}\Delta^{as}\Delta_{a}\right)\hskip 0.3cm \\ \nonumber
\tilde{\Delta}_b &=& \Delta_{b} - \frac{1}{E_{a}^2}\left((\Delta^{as})^2\Delta_{a} - (\xi^{as})^2\Delta_{a} + 2\xi^{as}\Delta^{as}\xi_{a}\right). 
\end{eqnarray}

\end{appendix}

\end{document}